\documentclass[12pt]{article}
\usepackage{amsmath}
\usepackage[round]{natbib}
\usepackage[english]{babel}
\usepackage[utf8]{inputenc}
\usepackage{accents}
\usepackage{graphicx}
\usepackage{rotating}
\usepackage{pdflscape}
\usepackage{float}
\usepackage[bottom]{footmisc}
\usepackage{ragged2e}

\usepackage{epsfig}
\usepackage{setspace} 
\usepackage{array}
\usepackage{multirow}
\usepackage{color}
\usepackage{subcaption}
\newcolumntype{L}[1]{>{\raggedright\let\newline\\\arraybackslash\hspace{0pt}}m{#1}}
\newcolumntype{C}[1]{>{\centering\let\newline\\\arraybackslash\hspace{0pt}}m{#1}}
\newcolumntype{R}[1]{>{\raggedleft\let\newline\\\arraybackslash\hspace{0pt}}m{#1}}

\newtheorem{thm}{Theorem}[subsection]

\DeclareMathOperator*{\argmax}{arg\,max}

\begin{document}

\title{A class of bootstrap based residuals for compositional data}

\author{Gustavo H. A. Pereira$^{1,2}$ \and Jianwen Cai$^2$}
\date{}

\maketitle

\footnotetext[1]{Department of Statistics, Federal University of São Carlos, Rod. Washington Luís, km 235 - SP-310 - São Carlos, CEP 13565-905,  Brazil. Email: gpereira@ufscar.br}
\footnotetext[2]{Department of Biostatistics, Gillings School of Global Public Health, University of North Carolina at Chapel Hill, Chapel Hill, NC, USA.}


\begin{abstract}

Regression models for compositional data are common in several areas of knowledge. As in other classes of regression models, it is desirable to perform diagnostic analysis in these models using residuals that are approximately standard normally distributed. However, for regression models for compositional data, there has not been any multivariate residual that meets this requirement. In this work, we introduce a class of asymptotically standard normally distributed residuals for compositional data based on bootstrap. Monte Carlo
simulation studies indicate that the distributions of the residuals of this class are well approximated by the standard normal distribution in small samples. An application to simulated data also suggests that one of the residuals of the proposed class is better to identify model misspecification than its competitors. Finally, the usefulness of the best residual of the proposed class is illustrated through an application on sleep stages. The class of residuals proposed here can also be used in other classes of multivariate regression models.

\end{abstract}

\vspace{0mm} \noindent {\textbf{Keywords}}: Bootstrap, compositional data, diagnostic analysis, quantile residual.

\vspace{3mm}

\section{Introduction}
\label{sec:introd}
\justify
\qquad 
Compositional data are proportions of disjoint categories
adding to one \citep{hijazi2009modelling}. They are common in several areas such as medicine \citep{gueorguieva2008dirichlet}, marketing \citep{morais2018using}, ecology \citep{tsilimigras2016compositional}, genetics \citep{fernandes2014unifying}, and geosciences \citep{filzmoser2012interpretation}. This kind of data can be modeled in different ways, but it is common to model them using the Dirichlet regression \citep{hijazi2009modelling} that can be viewed as a generalization of beta regression
\citep{ferrari2004beta}. An overview of advances in the analysis of compositional data can be found in \citet{alenazi2021review}.

When a regression model is fitted, it is important to check model adequacy and to identify outliers and influential observations. To reach these goals, it is of interest to use residuals whose distribution is well approximated by the standard normal distribution. When the distribution of a residual is not well approximated by the standard normal distribution, it is hard to interpret residuals plots. In these cases, it is not unusual that they are also nonidentically
distributed and well-fitted models may be discarded (see Application 3 of \citet{pereira2019quantile}).

The quantile residual \citep{dunn1996randomized} is simple and can be used in many regression models. It is asymptotically standard normally distributed if the parameters are consistently estimated. In addition, several papers showed that its distribution in small samples is  well approximated by the standard normal distribution for different regression models \citep{pereira2019quantile,lemonte2019residuals,feng2020comparison}.
Other works proposed modifications in the quantile residual to make it more suitable for some kinds of response variables \citep{scudilio2020adjusted,pereira2020class,andrade2022circular}.

For compositional data, the quantile residual can be calculated in two different ways. 
One can obtain the univariate or marginal quantile residual calculating it for each component of each observation \citep{hijazi2006residuals}. The problem with this approach is that it is hard to use a univariate residual for identifying outliers and for checking model adequacy in the analysis of compositional data, because each observation has several residuals. The other option is to calculate a single multivariate quantile residual for each observation using the distribution function assumed for the compositional response variable. However, the multivariate quantile residual may not be a good  measure of the discrepancy between the observed and the fitted value of the model. 


Several other residuals were proposed for compositional data.  \citet{gueorguieva2008dirichlet} describes three univariate residuals for compositional data, but they have the same drawback of the marginal quantile residual. \citet{gueorguieva2008dirichlet} also proposed two multivariate residuals. They are known as composite residuals, because they are a function of univariate residuals. However, these residuals assume only positive values and so their distribution cannot be well approximated by the standard normal distribution.


In this work, we propose a class of multivariate bootstrap based residuals for compositional data and prove that they are asymptotically standard normally distributed. Simulation studies also suggest that their distribution are well approximated by the standard normal distribution in small samples. This class of residuals use bootstrap to estimate the distribution function of a multivariate residual and apply the quantile function of the standard normal distribution for obtaining an asymptotically standard normally distributed residual. 

The remainder of this paper is organized as follows. Section \ref{sec:prop} presents the regression model considered in this work and introduces a  class of multivariate bootstrap based residuals for compositional data. In Section \ref{sec:sim}, Monte Carlo simulation studies are performed to study the properties of four residuals in the proposed class. An application to real data and to simulated data are presented in Section \ref{sec:appl}. Concluding remarks are provided in Section \ref{sec:concl}.

\section{A class of bootstrap based residuals}
\label{sec:prop}
\justify
\qquad 
The class of residuals proposed in this section  can be used in any parametric regression model for compositional response variable and independent observations. In the simulation studies and in the applications of our work, we consider the Dirichlet regression model, because it is the most used parametric model for compositional data. Some recent contributions related to this model include \citet{morais2018using}, \citet{ross2020tracking}, \citet{hanretty2021forecasting}, and \citet{melo2022higher}.  The Dirichlet regression model has been widely used in applications of several areas. See some recent applications of the model, for example, in \citet{vieira2019principled}, \citet{zhao2019additive}, \citet{chen2020associations}, \citet{breuninger2021associations}, and \citet{reichert2022age}.

\subsection{Dirichlet regression model}
\justify
\qquad 
There are two parameterizations of the Dirichlet distribution and they are known as common and alternative parameterization \citep{morais2018using}. In the former we have $k$ shape parameters, where $k$ is the number of components of the compositional random variable. Here we consider the alternative parametrization in which the parameters are more interpretable, and so it is more suitable for fitting regression models.

Let $w$ be a continuous compositional random variable, i.e., $w = (w_{1},w_{2},\hdots,w_{k})^\top$, $0 < w_{j} < 1$, $\sum_{j=1}^{k}w_{j}=1$. The compositional random variable $w$ is Dirichlet distributed with parameters $(\mu_1,\mu_2,\hdots,\mu_k,\phi)^\top, \mu_1= 1 - \sum_{j=2}^k \mu_j$, $0 < \mu_j < 1, \forall j \in \{1,2,\hdots,k\}$, and $\phi > 0$, if its probability density function is given by \citep{da2015bayesian}

\begin{equation}
\label{eq:pdf_dirich}
f(w_1,\hdots,w_k;\mu_1,\hdots,\mu_k,\phi)= \frac{\Gamma(\phi)}{\prod_{j=1}^k\Gamma(\phi\mu_j)}
\prod_{j=1}^kw_j^{\phi\mu_j-1}.
\end{equation}

In this parameterization, $\text{E}(w_j) = \mu_j$ and $\text{Var}(w_j) = \frac{\mu_j(1-\mu_j)}{1+\phi}$. Therefore the parameter $\mu_j$ is the mean of the component $j$ of the compositional random variable and $\phi$ is a precision parameter.

Let $y_1,y_2,\hdots,y_n, y_i = (y_{i1},y_{i2},\hdots, y_{ik})^\top$ be Dirichlet distributed random variables as defined in (\ref{eq:pdf_dirich}). The systematic components of the Dirichlet logistic regression model are given by 
\begin{equation}
\label{eq:reg_dirich}
\begin{array}{lll}
\log\left(\frac{\mu_{ij}}{\mu_{i1}}\right) & = & x_{ij1}\beta_{j1} + x_{ij2}\beta_{j2} + \hdots + x_{ijp_j}\beta_{jp_j}, \, 2 \leq j \leq k, \\
\log(\phi_i) & = & d_{i1}\gamma_{1} + d_{i2}\gamma_{2} + \hdots +
 d_{ip_\phi}\gamma_{p_\phi},
 \end{array}
\end{equation}
where ($x_{ij1},x_{ij2},\hdots,x_{ijp_j}, d_{i1},d_{i2},\hdots, d_{ip_\phi})^\top$ are constants that represent the values of the explanatory variables and 
$(\beta_{j1},\beta_{j2},\hdots,\beta_{jp_j},\gamma_{1},\gamma_{2},\hdots,\gamma_{p_\phi})^\top$ are unknown parameters. 
 Note that model (\ref{eq:reg_dirich}) uses the logit link function as is used in the multinomial logistic regression \citep{hosmer2013applied}. As in regression models with multinomial response \citep{das2014generalized,li2022alternate}, other link functions can be used in regression models with Dirichlet response. However, it is convenient to use the logit link function for the equations related to the mean vector of the response variable and the logarithmic link function for the dispersion parameter, because they lead to interpretable parameters. 

The parameters of model (\ref{eq:reg_dirich}) can be estimated by maximum likelihood and the model can be fitted using the package DirichletReg \citep{maier2014dirichletreg} of software R. We used this package in the simulation studies and in the applications presented in Sections \ref{sec:sim} and \ref{sec:appl}, respectively.

\subsection{A class of residuals for compositional data}



\justify
\qquad 
For regression models for univariate responses and independent observations, the quantile residual applies the inverse of the distribution function of the standard normal distribution on the distribution function of the response variable to obtain an asymptotically standard normally distributed residual. If we do the same for compositional data, using the multivariate distribution function of the response variable, we obtain a residual that is not necessarily a good  measure of the discrepancy between an observed and a fitted value of the model. This fact is a consequence of the multivariate nature of the compositional response variable. 

To obtain an asymptotically standard normally distributed residual that is also a good  measure of the discrepancy between an observed and a fitted value of the model, we modify the quantile residual. We change the distribution function of the response variable by the distribution function of a multivariate residual for compositional data. As the distribution of the multivariate residuals for compositional data are unknown, we use bootstrap \citep{efron1994introduction} to estimate the mentioned distribution function. Therefore, we propose a class of residuals that can be obtained using the following steps.
\begin{enumerate}
    \item For each observation, calculate $l_i=h_ir_i$, where $h_i$ is 1 or -1 (a sign function) and $r_i$ is a multivariate residual for compositional data. In the end of this section, we suggest two options for $h_i$ and $r_i$. 
    \item Generate $B$ bootstrap samples of $y=(y_1,y_2,\hdots,y_n)^\top$ using the fitted model and denote each of them as $y^{\{b\}}=(y_1^{\{b\}},y_2^{\{b\}},\hdots,y_n^{\{b\}})^\top$.
    \item For each bootstrap sample, fit the compositional regression model used for the original data considering the same covariates of the original model.
    \item For each bootstrap sample, obtain $l_i^{\{b\}}=h_i^{\{b\}}r_i^{\{b\}}$, where $h_i^{\{b\}}$ and $r_i^{\{b\}}$ are $h_i$ and $r_i$, respectively, for the $b{\text{th}}$ bootstrap sample.
    \item For each observation, obtain $a_i = \# \{l_i^{\{b\}} < l_i\}$.
    \item For each observation, randomly generate $u_i$ from a uniform distribution in the interval $(a_i/(B+1),(a_i+1)/(B+1))$.
    \item A residual of the proposed class is given by 
    $s_i = \Phi^{-1}(u_i)$, where $\Phi$ is the cumulative distribution function of the standard normal distribution.
\end{enumerate}

    
    Several observations related to the described procedure are necessary. The first ones refer to step 1. As we can use as $r_i$ any multivariate residual for compositional data and we can use different sign functions $h_i$, $s_i$ is a class of residuals, because for each different kind of $r_i$ and $h_i$, we have a different kind of $s_i$. 
    
    In compositional regression models, to define a  multivariate composite residual, usually a univariate residual is chosen and the multivariate residual is defined as the sum of the squared univariate residuals, across the $k$ components of each observation \citep{gueorguieva2008dirichlet}. We use this way of obtaining a multivariate residual, and also an alternative way in which we replace the quadratic function by the absolute value function. For the univariate residual, we consider the quantile residual, also known as pseudo residual in compositional data \citep{hijazi2006residuals}, because it is suggested as the best option to perform diagnostic analysis when the support of the response variable is the unit interval \citep{pereira2019quantile,lemonte2019residuals}. The univariate quantile residual for the Dirichlet regression model is defined as $q_{ij} = \Phi^{-1}[F(y_{ij};\hat{\mu}_{ij},\hat{\phi_i})]$, where $\Phi(.)$ and $F(.)$ are the distribution function of the standard normal distribution and of the beta distribution, respectively, and $\hat{\mu}_{ij}$ and $\hat{\phi}_i$ are the maximum likelihood estimator of $\mu_{ij}$ and $\phi_i$, respectively. We denote the multivariate residual $r_i$ as $r_i^{\{a\}}$ and $r_i^{\{q\}}$, when the absolute value function and the quadratic function are used to calculate $r_i$, respectively.

    Another observation is related to the sign function. It is important to use a sign function $h_i$, because the possible choices for $r_i$ assume only positive values and the best fitted observations have the lowest values of $r_i$. Therefore, if the sign function was not used, $s_i$ would have a high absolute value for the best fitted observations, because of the usage of the function $\Phi(.)$ in the step 7 of the procedure to obtain the residual. This is not an adequate behavior for a residual.
    
    We consider in this work two different definitions for the sign function $h_i$. For both definitions, we use an auxiliary random variable $m_i$ and $h_i = \text{sign}\{q_{im_i}\}$. In the first, for each observation, we consider the sign of the univariate quantile residual from the component that has the largest distance between the observed and the fitted value (largest absolute value of the univariate quantile residual). Therefore, in the first definition of $h_i$ denoted by $h_i^{\{1\}}$, $m_i^{\{1\}} = \underset{j \in \{1,2,\hdots,k\}}{\argmax} |q_{ij}|$. In the other definition, for all observations, we consider the sign of the univariate quantile residual from the component that has the largest average value of the response variable. Consequently, in this definition, $m_i^{\{2\}} = \underset{j \in \{1,2,\hdots,k\}}{\argmax} \sum_{i=1}^{n}y_{ij}/n$.
    
    The last observation is related to steps 6 and 7. It is important to consider $s_i = \Phi^{-1}(u_i)$ and not obtaining $s_i$ as a direct function of $a_i$ for two reasons. If we use $s_i$ as a direct function of $a_i$, we could not obtain $s_i$ when $a_i$ equals to 0 or 1. The second reason is to make $s_i$ continuous and allow it to converge in distribution to the standard normal distribution using a finite $B$. Note that, as we usually use a large $B$, the value of $s_i$ is not considerably affected by the randomization made in step 6. This strategy is similar to that used to obtain the randomized quantile residual for discrete data \citep{dunn1996randomized}. 
    
    Considering that we suggested two possible definitions for $r_i$ and for $h_i$, we consider in this work four different residuals in the proposed class: $s_i^{\{a1\}}$, $s_i^{\{q1\}}$, $s_i^{\{a2\}}$, and $s_i^{\{q2\}}$. Table \ref{ta:resid} summarizes the features of this four residuals.

 \begin{table} [h]   
 \caption {Features of the residuals of the proposed class.} 
     \vspace{-0.1cm} 
    \begin{center}
\begin{tabular} {ccc}  
\hline
Residual & Multivariate function used & Sign function \\
\hline
$s_i^{\{a1\}}$ & absolute value & 
$h_i = \text{sign}\{q_{im_i^{\{1\}}}\}$, 
$m_i^{\{1\}} = \underset{j \in \{1,2,\hdots,k\}}{\argmax} |q_{ij}|$ 
\\
$s_i^{\{q1\}}$ & quadratic & 
$h_i = \text{sign}\{q_{im_i^{\{1\}}}\}$, 
$m_i^{\{1\}} = \underset{j \in \{1,2,\hdots,k\}}{\argmax} |q_{ij}|$ 
\\
$s_i^{\{a2\}}$ & absolute value & 
$h_i = \text{sign}\{q_{im_i^{\{2\}}}\}$, 
$m_i^{\{2\}} = \underset{j \in \{1,2,\hdots,k\}}{\argmax} \sum_{i=1}^{n}y_{ij}/n$ 
\\
$s_i^{\{q2\}}$ & quadratic & 
$h_i = \text{sign}\{q_{im_i^{\{2\}}}\}$, 
$m_i^{\{2\}} = \underset{j \in \{1,2,\hdots,k\}}{\argmax} \sum_{i=1}^{n}y_{ij}/n$ 
\\

\hline
    \end{tabular}
\end{center}
    \label{ta:resid}
\end{table} 


As mentioned before, the following theorem holds for the residuals of the proposed class. It is proved in Appendix.
    
\begin{thm}
\label{teo:asympt}
Let $y_1, ..., y_n$ be compositional random variables, where each $y_i$ has vector of parameters $\tau_i$. If  $\tau_i$ is consistently estimated, under the true model,
$$
 s_i \xrightarrow{D} N(0;1).
$$
\end{thm}
    
    Theorem \ref{teo:asympt} guarantees that the proposed residuals are approximately standard normally distributed for large sample sizes. However, it is important to study if this approximation can be made in small samples and to evaluate their capacity to detect model misspecification. We perform these tasks in Sections \ref{sec:sim} and \ref{sec:applsim}, respectively.


\subsection{Other residuals for compositional data}
\justify
\qquad
The best known multivariate residual for compositional data is the composite residual proposed by \citet{gueorguieva2008dirichlet}. It was used for example by \citet{yoo2022guideline} and can be calculated in the package DirichletReg of the software R. This multivariate residual is a sum of the squared univariate Pearson residuals across the $k$ components of each observation. In the parameterization of the Dirichlet distribution used in this work, the composite residual is given by
\begin{equation}
\label{eq:res_comp}
r_i^{\{q1\}} = \sum_{j=1}^k \left(\frac{y_{ij} - \hat{\mu}_{ij}}{\sqrt{\hat{\phi}_i}}\right)^2.
\end{equation}

We also consider a residual similar to (\ref{eq:res_comp}), but using the quantile residual as the univariate residual. The composite quantile residual is given by
\begin{equation}
\label{eq:res_compq}
r_i^{\{q2\}} = \sum_{j=1}^k q_{ij}^2.
\end{equation}

Residual (\ref{eq:res_comp}) and  (\ref{eq:res_compq}) assume only positive values and so they are not approximately standard normally distributed. 
    
\section{Simulation studies}
\label{sec:sim}
\justify
\qquad
Monte Carlo simulation studies were performed to study if the distribution of the proposed residuals can be well approximated by the standard normal distribution in small samples. For all of them, we considered two values for the sample size (20 and 50), $k=3$, $B = 1000$ and 2000
Monte Carlo replicates. In each of the scenarios evaluated, we generated two covariates, $x_{i22}=x_{i32}=d_{i2}$, $x_{i23}=x_{i33}=d_{i3}$ and $x_{i21}=x_{i31}=d_{i1}=1$ for all $i$. The predictor variables were kept fixed across the 2000 replicates. All simulations were performed using the software R and the package DirichletReg. Given the goal of these simulation studies, we did not include residuals (\ref{eq:res_comp}) and (\ref{eq:res_compq}) in these analysis.

In scenario 1a, the covariates were generated from the standard uniform distribution and the value of the parameters were chosen in a way that the mean of $\mu_{i1}$, $\mu_{i2}$, and $\mu_{i3}$ across the $n$
observations are the same and the variance of the response variable is
high ($\phi$ close to 20). In this scenario, $\beta_{21}=-0.3,\beta_{22}=1.0,\beta_{23}=-0.5,\beta_{31}=-0.3,\beta_{32}=-0.5,\beta_{33}=1.0,\gamma_1=3.0,$ and $\gamma_2=\gamma_3=0$. 

In scenarios 2a to 5a, we changed a single feature of the first one. In scenarios 2a and 3a, we changed the value of some parameters to make the mean of the components of the response variable far and very far from each other, respectively. In scenario 4a, we considered that $\phi$ is a function of the covariates using $\gamma_2=0.5$ and $\gamma_3=-0.5$. In the following scenario, the first and second covariates were generated from the Bernoulli distribution with parameter 0.5 and Gamma distribution with parameters 3 and 6 (mean and variance as the standard uniform distribution), respectively. Scenarios 1b to 5b are similar to scenarios 1a to 5a, respectively, with $\gamma_1=4.6$ ($\phi$ close to 100 that corresponds to a low variance of the response variable). Table \ref{ta:scen} summarizes the considered scenarios.

 \begin{table} [h]   
 \caption {Description of the scenarios.} 
     \vspace{-0.1cm} 
    \begin{center}
\begin{tabular} {cccccccc}  
\hline
Scenarios & & \multicolumn{3}{c}{Mean of $(\mu_1,\mu_2,\mu_3)$} & & Covariates & $\phi$ \\
\cline{1-1} \cline{3-5} \cline{7-8}
1a, 1b & & 0.333 & 0.333 & 0.333 & & Uniform & Constant \\
2a, 2b & & 0.290 & 0.151 & 0.559 & & Uniform & Constant \\
3a, 3b & & 0.308 & 0.049 & 0.643 & & Uniform & Constant \\
4a, 4b & & 0.333 & 0.333 & 0.333 & & Uniform & Variable \\
5a, 5b & & 0.333 & 0.333 & 0.333 & & Bernoulli/Gamma & Constant \\
\hline
    \end{tabular}
\end{center}
    \label{ta:scen}
\end{table}

Table \ref{ta:scen1} presents the simulation results for scenario 1a and $n=20$. The table presents the sample mean, variance, skewness, and kurtosis coefficients for $s_i^{\{a1\}}$, $s_i^{\{q1\}}$, $s_i^{\{a2\}}$, and $s_i^{\{q2\}}$ for each of the $n$ observations across the $2000$ replicates. The table also contains the value of the Anderson–Darling (AD) statistic \citep{stephens1986tests} that can be used to test whether each residual is standard normally distributed.
The mean and the variance of residuals are very close to zero and one, respectively, for all observations and proposed residuals. The skewness coefficient is also very close to zero for almost all observations and residuals and not so far from zero for the remaining observations. The kurtosis coefficients of the residuals are farther for the corresponding value of the standard normal distribution than the other measures but there are no observations that have kurtosis coefficient very far from 3 for any residual. As a consequence of these results, the values of the AD statistics are all small and the distribution of the four proposed residual seems to be well approximated by the standard normal distribution in this scenario with $n=20$. The quality of this approximation seems to be similar for the four residuals since the mean of the AD statistics are very close to each other. 

Table \ref{ta:sum_scen} presents the average values for the mean, variance, skewness, kurtosis coefficient, and AD Statistics for all proposed residuals in the ten studied scenarios with $n=20$. The distribution of the four proposed residuals seems to be well approximated by the standard normal distribution in all scenarios when $n=20$, since the average values of all measures for the four proposed residuals are very close to theoretical values of this distribution. The average values of the AD statistics are small for the four residuals in the ten scenarios suggesting that this approximation is not affected by the value of the parameters related to the mean of the response variable, nor by the variance of the response and nor by the distribution of the covariates.  The quality of the standard normality approximation seems to be similar for the four residuals in all scenarios. Results for $n = 50$ are similar and not included here for brevity.

\section{Applications}
\label{sec:appl}
\qquad

We present two applications. The first application uses simulated data and its goal is to compare the ability of the residuals to detect model misspecification. We also analyzed data on sleep stages to illustrate the usefulness of the best residual based on the simulation results in the first application.

\subsection{Application to simulated data}
\label{sec:applsim}
\qquad

For evaluating if a parametric regression model properly fits a certain response variable, a normal probability plot with simulated envelope \citep{atkinson1981two} is commonly used. If all residuals are inside the envelope or if only few points are outside the envelope and they are all close to one of the bounds of the envelope, there is no evidence that the model is inadequate.

\begin{landscape}

 \begin{table} [h]   
 \caption {Simulation results Scenario 1a - $n=20$.} 
     \vspace{-0.3cm} 
    \begin{center}
\tabcolsep=0.1cm 
\scalebox{0.93}{
\begin{tabular} {rrrrrrrrrrrrrrrrrrrrrrrrr}      
    \hline
 $i$ & \multicolumn{4} {c} {Mean} & & \multicolumn{4} {c} {Variance} & & \multicolumn{4} {c} {Skewness} & & \multicolumn{4} {c} {Kurtosis} & & \multicolumn{4} {c} {A-D Statistic} \\
\cline{2-5} \cline{7-10} \cline{12-15} \cline{17-20} \cline{22-25}    
	&	$s_i^{\{a1\}}$	&	$s_i^{\{q1\}}$	&	$s_i^{\{a2\}}$	&	$s_i^{\{q2\}}$	&	&	$s_i^{\{a1\}}$	&	$s_i^{\{q1\}}$	&	$s_i^{\{a2\}}$	&	$s_i^{\{q2\}}$	&	&	$s_i^{\{a1\}}$	&	$s_i^{\{q1\}}$	&	$s_i^{\{a2\}}$	&	$s_i^{\{q2\}}$	&	&	$s_i^{\{a1\}}$	&	$s_i^{\{q1\}}$	&	$s_i^{\{a2\}}$	&	$s_i^{\{q2\}}$	&	&	$s_i^{\{a1\}}$	&	$s_i^{\{q1\}}$	&	$s_i^{\{a2\}}$	&	$s_i^{\{q2\}}$	\\
\hline																																												
1	&	0.01	&	0.01	&	0.01	&	0.01	&	&	0.94	&	0.94	&	0.95	&	0.94	&	&	0.03	&	0.06	&	0.00	&	-0.03	&	&	2.96	&	3.09	&	3.02	&	3.01	&	&	0.88	&	0.94	&	0.76	&	0.82	\\
2	&	-0.01	&	-0.02	&	0.02	&	0.02	&	&	1.06	&	1.06	&	1.06	&	1.06	&	&	0.13	&	0.10	&	-0.01	&	-0.01	&	&	2.96	&	2.88	&	2.98	&	2.93	&	&	1.43	&	1.45	&	1.23	&	1.21	\\
3	&	0.01	&	0.01	&	0.04	&	0.04	&	&	1.05	&	1.05	&	1.06	&	1.06	&	&	-0.07	&	-0.07	&	-0.02	&	0.01	&	&	3.09	&	3.16	&	3.20	&	3.22	&	&	0.72	&	0.83	&	1.63	&	1.77	\\
4	&	-0.02	&	-0.02	&	-0.01	&	-0.01	&	&	0.97	&	0.98	&	0.97	&	0.98	&	&	-0.06	&	-0.04	&	0.03	&	0.03	&	&	2.91	&	2.90	&	2.90	&	2.90	&	&	0.52	&	0.49	&	0.46	&	0.40	\\
5	&	0.02	&	0.02	&	-0.01	&	-0.01	&	&	0.98	&	0.96	&	0.98	&	0.97	&	&	0.11	&	0.10	&	-0.04	&	-0.03	&	&	3.09	&	2.99	&	3.13	&	3.05	&	&	1.19	&	1.16	&	0.74	&	0.75	\\
6	&	0.00	&	0.00	&	-0.01	&	-0.01	&	&	0.99	&	0.99	&	0.99	&	0.99	&	&	0.05	&	0.05	&	-0.03	&	-0.04	&	&	3.31	&	3.37	&	3.40	&	3.40	&	&	1.09	&	1.21	&	1.14	&	1.25	\\
7	&	0.01	&	0.01	&	-0.01	&	-0.01	&	&	1.04	&	1.04	&	1.04	&	1.04	&	&	0.04	&	0.00	&	-0.17	&	-0.14	&	&	3.13	&	3.17	&	3.10	&	3.09	&	&	0.67	&	0.75	&	0.98	&	0.92	\\
8	&	0.00	&	0.00	&	-0.03	&	-0.04	&	&	1.02	&	1.02	&	1.02	&	1.02	&	&	0.03	&	0.03	&	0.01	&	-0.03	&	&	3.02	&	2.99	&	3.03	&	2.98	&	&	0.42	&	0.39	&	1.13	&	1.30	\\
9	&	0.03	&	0.04	&	-0.01	&	-0.01	&	&	1.02	&	1.03	&	1.03	&	1.03	&	&	0.09	&	0.11	&	0.01	&	0.00	&	&	3.06	&	3.11	&	3.14	&	3.08	&	&	1.17	&	1.20	&	0.47	&	0.40	\\
10	&	0.00	&	-0.01	&	0.02	&	0.02	&	&	1.02	&	1.02	&	1.02	&	1.02	&	&	0.08	&	0.06	&	-0.06	&	-0.06	&	&	2.99	&	3.02	&	3.04	&	3.07	&	&	0.61	&	0.58	&	0.92	&	0.92	\\
11	&	-0.02	&	-0.03	&	0.01	&	0.01	&	&	0.95	&	0.95	&	0.95	&	0.95	&	&	0.05	&	0.02	&	-0.07	&	-0.06	&	&	3.00	&	3.03	&	3.06	&	3.08	&	&	1.37	&	1.38	&	0.86	&	0.89	\\
12	&	-0.02	&	-0.02	&	0.02	&	0.02	&	&	0.99	&	0.99	&	0.99	&	0.99	&	&	0.00	&	-0.01	&	-0.04	&	-0.09	&	&	2.98	&	2.95	&	2.90	&	2.94	&	&	0.46	&	0.50	&	0.86	&	0.91	\\
13	&	0.01	&	0.00	&	0.02	&	0.02	&	&	1.02	&	1.03	&	1.03	&	1.03	&	&	0.01	&	-0.01	&	0.01	&	-0.01	&	&	2.94	&	2.96	&	2.92	&	2.94	&	&	0.71	&	0.63	&	0.88	&	0.86	\\
14	&	0.04	&	0.03	&	-0.03	&	-0.03	&	&	0.99	&	0.99	&	1.00	&	1.01	&	&	-0.03	&	-0.03	&	0.02	&	0.01	&	&	3.01	&	3.03	&	3.05	&	3.11	&	&	1.85	&	1.90	&	1.04	&	1.01	\\
15	&	0.03	&	0.03	&	-0.04	&	-0.04	&	&	1.00	&	1.00	&	1.01	&	1.00	&	&	0.08	&	0.07	&	-0.01	&	-0.02	&	&	2.91	&	2.89	&	3.02	&	2.95	&	&	0.95	&	0.91	&	1.39	&	1.44	\\
16	&	0.01	&	0.01	&	-0.01	&	-0.01	&	&	0.97	&	0.98	&	0.98	&	0.98	&	&	0.07	&	0.03	&	-0.03	&	-0.01	&	&	2.82	&	2.84	&	2.89	&	2.86	&	&	0.30	&	0.21	&	0.30	&	0.23	\\
17	&	-0.02	&	-0.03	&	0.02	&	0.01	&	&	0.99	&	0.98	&	0.98	&	0.98	&	&	0.13	&	0.09	&	0.03	&	0.00	&	&	3.26	&	3.01	&	3.02	&	3.03	&	&	1.13	&	1.08	&	0.81	&	0.79	\\
18	&	-0.02	&	-0.02	&	0.00	&	0.00	&	&	1.03	&	1.04	&	1.03	&	1.03	&	&	-0.05	&	-0.02	&	-0.04	&	-0.03	&	&	2.95	&	2.93	&	2.87	&	2.94	&	&	0.70	&	0.65	&	0.41	&	0.43	\\
19	&	0.02	&	0.02	&	0.00	&	0.00	&	&	0.94	&	0.95	&	0.96	&	0.96	&	&	0.01	&	-0.03	&	-0.05	&	-0.07	&	&	2.94	&	3.04	&	3.01	&	3.07	&	&	1.01	&	0.95	&	0.54	&	0.53	\\
20	&	0.00	&	0.00	&	-0.04	&	-0.04	&	&	1.02	&	1.02	&	1.02	&	1.02	&	&	0.07	&	0.05	&	-0.03	&	-0.02	&	&	2.90	&	2.89	&	2.97	&	2.94	&	&	0.67	&	0.62	&	1.76	&	1.95	\\
\hline																																													
Mean	&	0.00	&	0.00	&	0.00	&	0.00	&	&	1.00	&	1.00	&	1.00	&	1.00	&	&	0.04	&	0.03	&	-0.03	&	-0.03	&	&	3.01	&	3.01	&	3.03	&	3.03	&	&	0.89	&	0.89	&	0.92	&	0.94	\\
SD	&	0.02	&	0.02	&	0.02	&	0.02	&	&	0.03	&	0.04	&	0.03	&	0.03	&	&	0.06	&	0.05	&	0.05	&	0.04	&	&	0.12	&	0.13	&	0.12	&	0.12	&	&	0.39	&	0.41	&	0.39	&	0.45	\\

\hline
    \end{tabular}
}
\end{center}
    \label{ta:scen1}
\end{table}
\end{landscape}

 \begin{table} [h!]   
 \caption {Average distributional measures for the residuals - $n=20$.} 
     \vspace{-0.1cm} 
    \begin{center}
\begin{tabular} {ccrrrrrrrrr}      
    \hline
Scenarios & & \multicolumn{4} {c} {$\gamma_1=3.0$} & &   \multicolumn{4} {c} {$\gamma_1=4.6$}  \\
\cline{3-6} \cline{8-11}
	&		&	$s_i^{\{a1\}}$	&	$s_i^{\{q1\}}$	&	$s_i^{\{a2\}}$	&	$s_i^{\{q2\}}$	&	&	$s_i^{\{a1\}}$	&	$s_i^{\{q1\}}$	&	$s_i^{\{a2\}}$	&	$s_i^{\{q2\}}$	\\
\hline																				
	&	Mean	&	0.00	&	0.00	&	0.00	&	0.00	&	&	0.00	&	0.00	&	0.00	&	0.00	\\
	&	Variance	&	1.00	&	1.00	&	1.00	&	1.00	&	&	1.00	&	1.00	&	1.00	&	1.00	\\
1	&	Skewness	&	0.04	&	0.03	&	-0.03	&	-0.03	&	&	0.00	&	-0.01	&	0.01	&	0.00	\\
	&	Kurtosis	&	3.01	&	3.01	&	3.03	&	3.03	&	&	2.98	&	3.01	&	2.96	&	2.99	\\
	&	A-D Statistics	&	0.89	&	0.89	&	0.92	&	0.94	&	&	0.90	&	0.87	&	0.92	&	0.92	\\
\hline																				
	&	Mean	&	0.00	&	0.00	&	0.00	&	0.00	&	&	0.00	&	-0.01	&	0.00	&	0.00	\\
	&	Variance	&	1.00	&	1.00	&	1.00	&	1.00	&	&	1.00	&	1.00	&	1.00	&	1.00	\\
2	&	Skewness	&	0.01	&	0.00	&	0.01	&	0.01	&	&	0.00	&	0.00	&	-0.01	&	-0.01	\\
	&	Kurtosis	&	2.99	&	3.00	&	3.00	&	2.99	&	&	2.99	&	2.99	&	3.01	&	3.01	\\
	&	A-D Statistics	&	0.91	&	0.90	&	0.98	&	0.99	&	&	0.99	&	0.99	&	0.80	&	0.80	\\
\hline																				
	&	Mean	&	0.00	&	0.00	&	0.00	&	0.00	&	&	0.01	&	0.01	&	0.00	&	0.00	\\
	&	Variance	&	1.00	&	1.00	&	1.00	&	1.00	&	&	1.00	&	1.00	&	1.00	&	1.00	\\
3	&	Skewness	&	-0.01	&	-0.01	&	0.00	&	0.01	&	&	0.00	&	0.00	&	0.00	&	0.00	\\
	&	Kurtosis	&	2.99	&	3.00	&	2.97	&	2.97	&	&	3.01	&	3.02	&	2.99	&	3.01	\\
	&	A-D Statistics	&	1.24	&	1.27	&	0.81	&	0.79	&	&	0.84	&	0.83	&	0.86	&	0.81	\\
\hline																				
	&	Mean	&	0.01	&	0.01	&	0.00	&	0.00	&	&	0.01	&	0.01	&	0.00	&	0.00	\\
	&	Variance	&	0.99	&	1.00	&	1.00	&	1.00	&	&	1.00	&	1.00	&	1.00	&	1.00	\\
4	&	Skewness	&	0.01	&	0.00	&	-0.01	&	0.00	&	&	0.00	&	0.00	&	0.00	&	0.00	\\
	&	Kurtosis	&	2.95	&	2.98	&	2.99	&	3.00	&	&	2.99	&	3.01	&	2.97	&	2.99	\\
	&	A-D Statistics	&	1.46	&	1.46	&	1.22	&	1.18	&	&	0.91	&	0.89	&	1.17	&	1.11	\\
\hline																				
	&	Mean	&	0.01	&	0.01	&	0.00	&	0.00	&	&	0.00	&	0.00	&	0.00	&	0.00	\\
	&	Variance	&	1.00	&	1.00	&	1.00	&	1.00	&	&	1.00	&	1.00	&	1.00	&	1.00	\\
5	&	Skewness	&	0.01	&	0.01	&	-0.01	&	-0.01	&	&	0.00	&	0.00	&	-0.01	&	-0.02	\\
	&	Kurtosis	&	3.00	&	3.01	&	3.03	&	3.01	&	&	2.97	&	2.97	&	2.97	&	2.97	\\
	&	A-D Statistics	&	0.65	&	0.66	&	0.90	&	0.91	&	&	1.08	&	1.05	&	0.88	&	0.85	\\

\hline																			
    \end{tabular}
\end{center}
    \label{ta:sum_scen}
\end{table}

There is no general rule in the literature to define how far a residual should be to the closest bound of the envelope to lead to rejection of the model. However, for a proper comparison of the ability of the residuals to detect model misspecification, it is important to define a rule for model rejection. Let $v$ be a constant for which, when the correct model is fitted, the probability of at least one residual is outside the envelope and is more than $v$ units away from the closest bound of the envelope is $5\%$. In practice, with a given data set, $v$ will depend on the sample size $n$  and the dispersion of the residuals.

Consider that $g$ datasets of size $n$ were generated from a certain regression model. Consider also that for each of these datasets, the correct model was fitted, a certain residual $r_i$ was calculated for the $n$ observations, and a normal probability plot with simulated envelope was made. Let $e_j$ be zero if all residuals of the $j$th datasets are inside the simulated envelope and be the distance from the residual that is farthest from the envelope to the closest bound of the envelope. Using this definition, a natural estimator of $v$ is given by $(e_{[0.95g]} + e_{[0.95g+1]})/2$, where $e_{[j]}$ denotes the $j$th lowest value of $e_j$ across the $g$ datasets.

To estimate $v$ for the residuals considered in this work, we generated 500 datasets based on model (\ref{eq:reg_dirich}) with $n=50$ and $k=3$. The two covariates were generated from the binomial distribution with parameters $n$ and 0.5 and the Dirichlet distributed response variable was generated according to the set 1 of Distributions presented on Table
\ref{ta:appl_sim}.

 \begin{table}[ht]   
 \caption {Probability distributions used in the application with simulated data.} 
     \vspace{-0.1cm} 
   \begin{center}
\begin{tabular} {cccc}      
    \hline
$x_{i22}$ & $x_{i23}$ & Set 1 of distributions & Set 2 of distributions \\
\hline
1 & 0 & Dirichlet$(0.80,0.15,0.05,e^{4.6})$ & Dirichlet$(0.40,0.35,0.25,e^{4.6})$ \\
0 & 1 & Dirichlet$(0.05,0.80,0.15,e^{4.6})$ & Dirichlet$(0.25,0.40,0.35,e^{4.6})$ \\
1 & 1 & Dirichlet$(0.15,0.05,0.80,e^{4.6})$ & Dirichlet$(0.35,0.25,0.40,e^{4.6})$ \\
0 & 0 & Dirichlet$(0.35,0.35,0.30,e^{4.6})$ & Dirichlet$(0.30,0.30,0.35,e^{4.6})$ \\
\hline																	    \end{tabular}
\end{center}
    \label{ta:appl_sim}
\end{table}

For each of the datasets, we fitted the correct model, calculated the residuals presented in Section \ref{sec:prop}, and made normal probability plots with simulated envelope. As the residuals of the proposed class have a similar variance as presented in Table \ref{ta:sum_scen}, we used the same $v$ for these four residuals. To estimate it, we used the plots for these four residuals and denoted its natural estimator by $v^{\{1\}}$. Residuals (\ref{eq:res_comp}) and (\ref{eq:res_compq}) are a sum of squared univariate residuals, and they have the same variances but are different from those from the proposed class. Therefore, we used the same $v$ for these two residuals and denoted its estimator as $v^{\{2\}}$.

\begin{figure}[ht]
\centering 
\includegraphics[width=12cm]{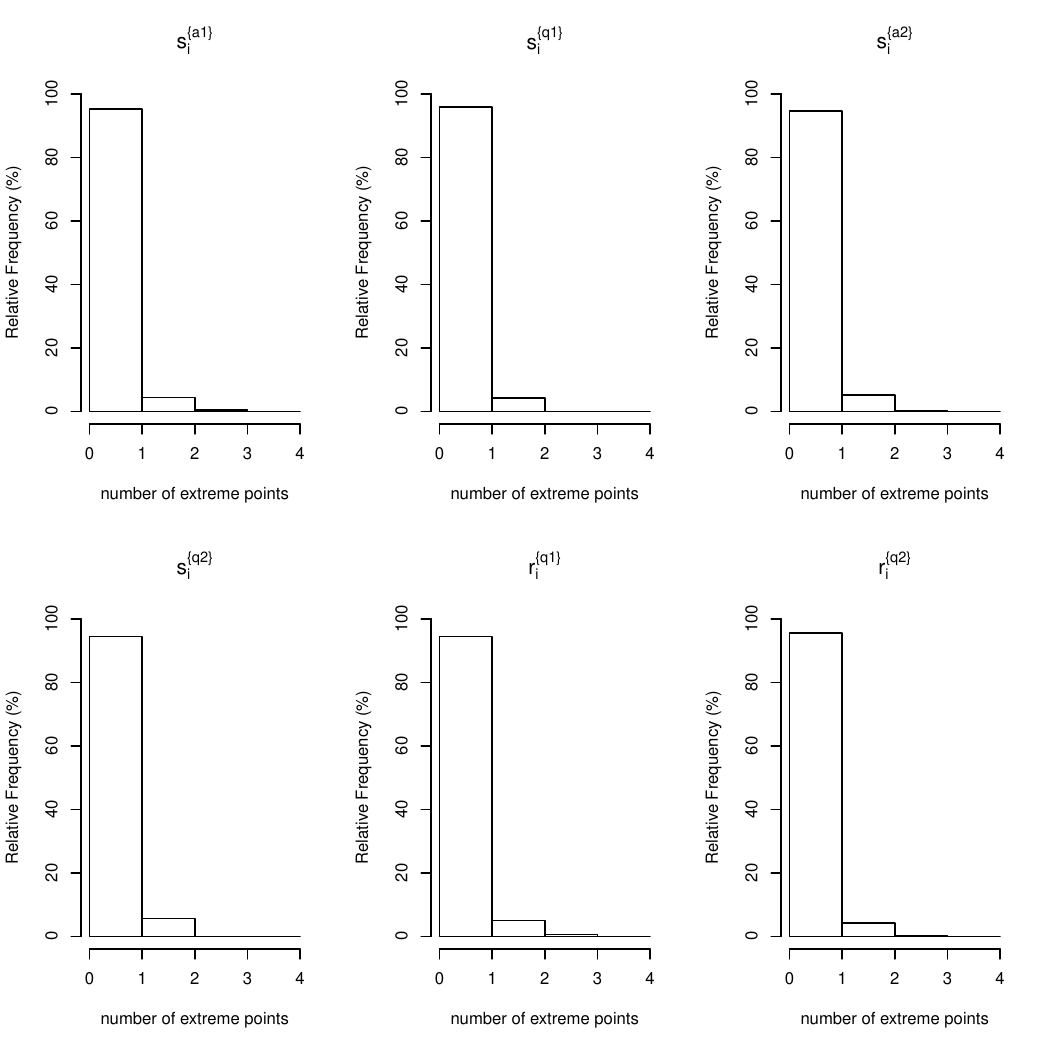} 
\caption{Histogram for the number of points outside the simulated envelope and $v^{\{j\}}$ units or more away from the closest envelope bounds - Correct model.}
\label{fig:sim_correct}
\end{figure}

Using the 500 generated datasets, Figure \ref{fig:sim_correct} presents a histogram for the number of points outside the simulated envelope and $v^{\{j\}}$ units or more away from the closest envelope bounds, where $j=1$ for the residuals of the proposed class and $j=2$ for $r_i^{\{q1\}}$ and $r_i^{\{q2\}}$. As expected, the plots are similar and, for the six residuals, about $95\%$ of datasets do not have any point outside the simulated envelope and $v^{\{j\}}$ units or more away from the closest envelope bounds.

To compare the ability of the residuals to detect model misspecification, we generated 500 additional datasets with $k=3$ and $n=50$. The covariates were generated in the same way as in the previous datasets, and the response variable was generated as a mixture of the set of Distributions 1 and 2  (Table \ref{ta:appl_sim}) with probabilities 0.7 and 0.3, respectively. For each of these datasets, we incorrectly fitted model (\ref{eq:reg_dirich}), calculated the residuals presented in Section \ref{sec:prop}, and made normal probability plots with simulated envelope.

\begin{figure}[ht]
\centering 
\includegraphics[width=12cm]{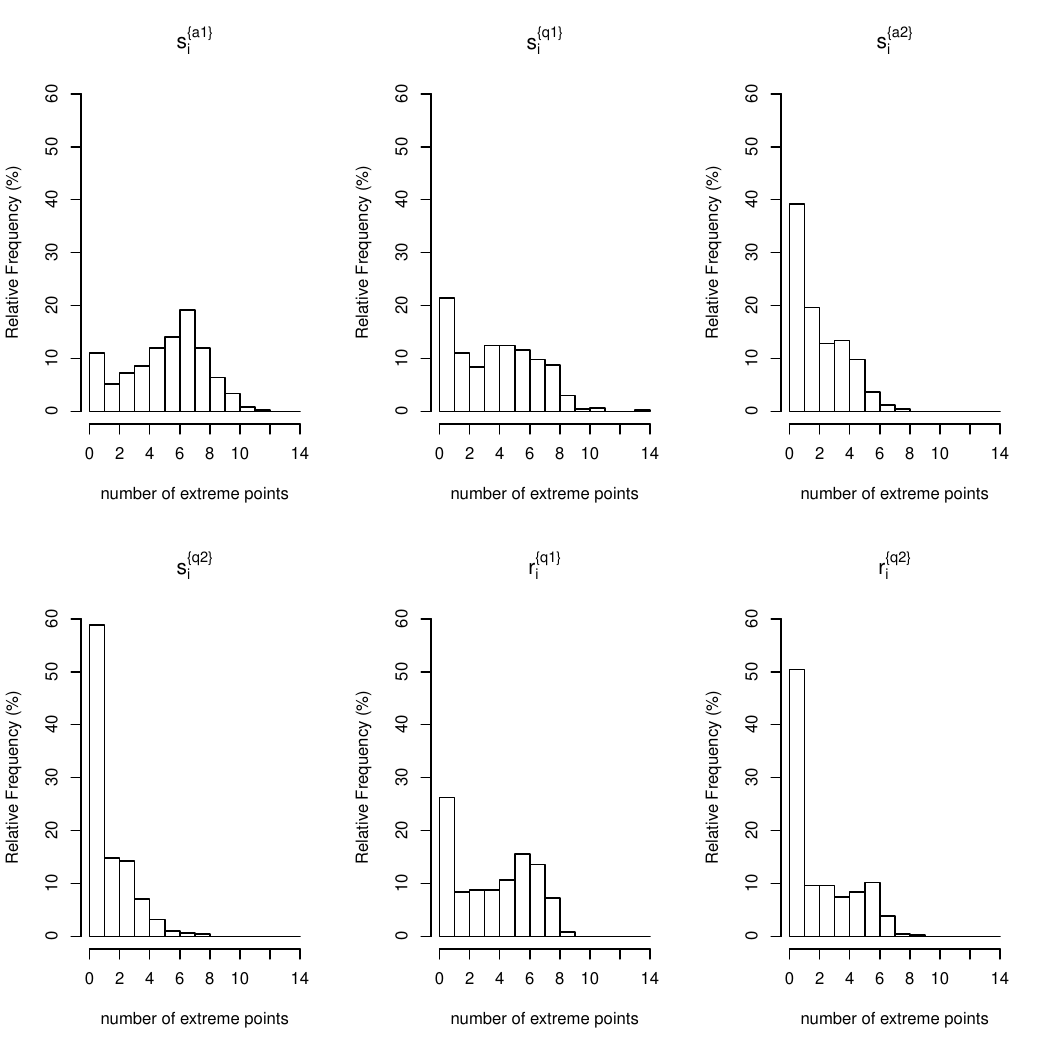} 
\caption{Histogram for the number of points outside the simulated envelope and $v^{\{j\}}$ units or more away from envelope bounds - Wrong model.}
\label{fig:sim_wrong}
\end{figure}

Figure \ref{fig:sim_wrong} presents a histogram for the number of points outside the simulated envelope and $v^{\{j\}}$ units or more away from the closest envelope bounds when the incorrect model was fitted. The residual $s_i^{\{a1\}}$ of the proposed class is the best to detect model misspecification since, for this residual, only $11\%$ of the datasets do not have any point outside the envelope and distant from the bounds. In addition, using $s_i^{\{a1\}}$, more than $70\%$ of the datasets have at least 3 points outside the envelope and far from the bounds. On the other hand, the other residuals could not detect model misspecification for more than $20\%$ of the datasets.  

\subsection{Application to data on sleep stages}
\label{sec:applsleep}

\qquad
The human sleep has four stages: N1, N2, N3 and REM. During the first two stages, the person is in a light sleep \citep{ritmala2015sleep}. The  stage N3 corresponds a deep sleep and it is when muscles and tissues are healed \citep{hussain2022quantitative}. Finally, the stage REM is characterized by rapid eye movements and fast breathing and this stage is responsible for optimal emotional processing \citep{tempesta2018sleep}.

Considering the functions of the last two stages, it is important that people spend a considerable percentage of their sleep on stages N3 and REM. Sleep is considered as within normal parameters when the percentage time spent on the stages N1, N2, N3, and REM are on the following ranges: $3-8\%$, $45-55\%$, $15-20\%$, $20-25\%$, respectively \citep{gonzalez2019analysis}. 

The proportion of time spent on each sleep stage is a compositional variable. However, papers usually study each of the stages separately considering  four independent proportions \citep{ohayon2017national,maski2021stability}. For the best of our knowledge, this paper is the first that considers the proportion of time spent on each sleep stage as a compositional variable.

\citet{veje2021sleep} collected sleep data from 42 adults, in which 22 were patients with Tick-borne encephalitis (TBE) and 20 controls. Their main goal was to investigate if the TBE affects the quality of the sleep measured by a questionnaire. Here, we use a Dirichlet regression model to study if the proportion of time spent on each sleep stages changes with the total sleep time (TST) and from patients with TBE to healthy adults. The sample means for the proportion of time spent on N1, N2, N3 and REM stages are 0.134, 0.532, 0.152, and 0.182, respectively. These sample means are close for TBE adults and control and the standard deviations are less than 0.1.

The Dirichlet regression model assumes that $P(y_{ij}=0) = 0$ for all $i$ and $j$. As a result, the model can not be fitted if there are one or more zeros in the response variable. When there are many zeros, a possible solution is to fit a zero adjusted Dirichlet regression model \citep{tsagris2018dirichlet}. However, the parameters of the discrete component of this model can not be well estimated when there are few zeros. In this case, it is better to transform data in a way that eliminates the zeros \citep{hijazi2011algorithm}. When there are very few zeros, it is reasonable to simply replacing the zeros by a small value and then normalizing the imputed compositions \citep{aitchison1986stat}.

Two patients of the data collected by \citet{veje2021sleep} do not spend time on the N3 stage in their monitored night of sleep. We replaced their zero values by 0.001 and normalized the proportions of these patients to sum 1. 

We fitted a Dirichlet regression model for the transformed data containing the mentioned covariates in all submodels. Using the log likelihood ratio test, we concluded that there are no evidences that the parameters of the proportion of time spent on each sleep stage changes from patients with TBE to healthy adults (p = 0.7469). However, the mean vector (p = 0.0019) and the precision parameter (p = 0.0030) of the response variable is a function of the total sleep time.

\begin{figure}[ht]
\centering 
\includegraphics[width=12cm]{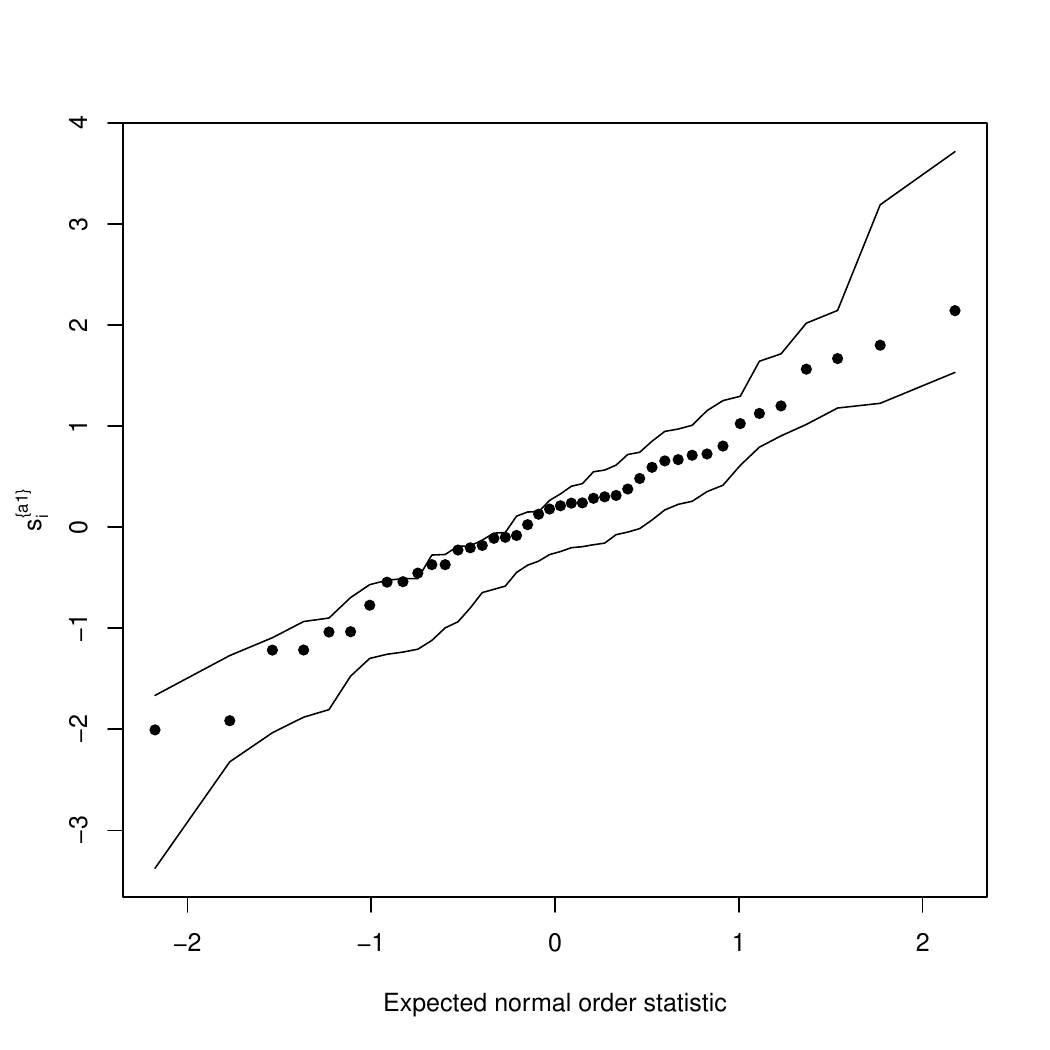} 
\caption{Normal probability plot with simulated envelope for $s_i^{\{a1\}}$ considering the final model for data on sleep stages}
\label{fig:envel}
\end{figure}

As the residual $s_i^{\{a1\}}$ performed better in the analysis with simulated data, we used it to conduct the diagnostic analysis. Figure \ref{fig:envel} presents a 
normal probability plot with simulated envelope for $s_i^{\{a1\}}$ considering the final model. The plot does not suggest model misspecification of the Dirichlet regression model for the proportion of time spent on each sleep stage with the total sleep time as a covariate of all submodels.


Table \ref{ta:appl_real} presents the parameter estimates and standard errors for the final model. Note that the estimates of the parameters associated with TST in the second and third submodels are positive. Therefore, on average, an increase in the total sleep time leads to a higher proportion of time spent on stages N3 and REM in relation to the N1 stage. These results suggest that adults have a better quality of sleep if they spend enough time on sleeping.

Calculating the exponential of the parameter estimates (last column of Table \ref{ta:appl_real}), it is estimated that for every hour increase in the total sleep
time, the ratio between the mean of the proportion of time spent on the N3 stage of the sleep and the mean of the proportion of time spent on the N1 stage increases by 40.3\%. Similarly,  it is estimated that for every hour increase in the total sleep time, the ratio between the mean of the proportion of time spent on the REM stage of the sleep and the mean of the proportion of time spent on the N1 stage increases by 23.8\%

 \begin{table}[ht]   
 \caption {Parameter estimates and standard errors in the model
for the proportion of time spent on each sleep stage.} 
     \vspace{-0.1cm} 
   \begin{center}
\begin{tabular} {ccrrr}      
\hline									
Submodel	&	Covariate	&	Estimate	&	Std. Error	&	Exp(estim)	\\
\hline									
$\mu_{N2}$	&	Intercept	&	0.5142	&	0.5949	&	1.672	\\
	&	TST	&	0.1314	&	0.0880	&	1.140	\\
\hline									
$\mu_{N3}$	&	Intercept	&	-2.2371	&	0.7087	&	0.107	\\
	&	TST	&	0.3386	&	0.1030	&	1.403	\\
\hline									
$\mu_{REM}$	&	Intercept	&	-1.0275	&	0.6688	&	0.358	\\
	&	TST	&	0.2136	&	0.0981	&	1.238	\\
\hline									
$\phi$	&	Intercept	&	0.8401	&	0.7433	&	2.317	\\
	&	TST	&	0.3400	&	0.1139	&	1.405	\\
\hline
\end{tabular}
\end{center}
    \label{ta:appl_real}
\end{table}

\section{Concluding remarks}
\label{sec:concl}
\qquad
In this work, we proposed a class of asymptotically standard normally distributed residuals for compositional data based on bootstrap. We investigated the properties of four residuals of the proposed class in small samples using Monte Carlo simulation studies and an application on simulated data considering the Dirichlet regression model. We also presented an application on the stages of the sleep that illustrates the usefulness of one of the residuals of the proposed class.

To facilitate the diagnostic analysis of a regression model, it is desirable to find a residual whose distribution is well approximated by the standard normal distribution. To fill in the gap in the literature for compositional data, we proposed a class of residuals and proved that the residuals of the proposed class are asymptotically standard normally distributed, and hence their distribution is close to the standard normal distribution when the sample size is large. 
In addition, simulation studies suggested that the distributions of the four investigated residual of the proposed class are very close to the standard normal distribution even when $n=20$ for all considered scenarios.

It is also important that a residual can detect model misspecification. An application using simulated data suggested that the residual of the proposed class denoted by $s_i^{\{a1\}}$ is good for identifying model misspecification and better than the other residuals considered in this paper. 

Another interesting feature of the proposed class of residuals is that their expressions do not depend on the assumed distribution of the response variable. The algebraic expression of the multivariate residuals based on the deviance residual, for example, have to be obtained for each different distribution assumed to the response variable. 

Besides the usage for diagnostic analysis in regression models for compositional data, the proposed class of residuals can also be used in other classes of multivariate regression models. There is no approximately standard normally distributed multivariate residual in the literature for multivariate regression models. Therefore, our class of residuals can be a useful tool for several classes of multivariate regression models. Future work can investigate the properties of residuals of the proposed class in different multivariate regression models, such as the multivariate normal regression model \citep{johnson2013applied}, the multivariate elliptical regression model \citep{lemonte2011multivariate}, and the multivariate generalized linear model \citep{schaid2019multivariate}.

\section{Acknowledgment}

This work was partially supported by São Paulo Research Foundation (FAPESP), grant number 2020/16334-9.


\singlespacing   

\bibliographystyle{elsarticle-harv}


\begin{thebibliography}{47}
\expandafter\ifx\csname natexlab\endcsname\relax\def\natexlab#1{#1}\fi
\providecommand{\url}[1]{\texttt{#1}}
\providecommand{\href}[2]{#2}
\providecommand{\path}[1]{#1}
\providecommand{\DOIprefix}{doi:}
\providecommand{\ArXivprefix}{arXiv:}
\providecommand{\URLprefix}{URL: }
\providecommand{\Pubmedprefix}{pmid:}
\providecommand{\doi}[1]{\href{http://dx.doi.org/#1}{\path{#1}}}
\providecommand{\Pubmed}[1]{\href{pmid:#1}{\path{#1}}}
\providecommand{\bibinfo}[2]{#2}
\ifx\xfnm\relax \def\xfnm[#1]{\unskip,\space#1}\fi
\bibitem[{Aitchison(1986)}]{aitchison1986stat}
\bibinfo{author}{Aitchison, J.}, \bibinfo{year}{1986}.
\newblock \bibinfo{title}{The Statistical Analysis of Compositional Data}.
\newblock \bibinfo{publisher}{Springer}.
\bibitem[{Alenazi(2023)}]{alenazi2021review}
\bibinfo{author}{Alenazi, A.}, \bibinfo{year}{2023}.
\newblock \bibinfo{title}{A review of compositional data analysis and recent advances}.
\newblock \bibinfo{journal}{Communications in Statistics-Theory and Methods} \bibinfo{volume}{52}, \bibinfo{pages}{5535--5567}.
\bibitem[{Andrade et~al.(2023)Andrade, Pereira and Artes}]{andrade2022circular}
\bibinfo{author}{Andrade, A.C.}, \bibinfo{author}{Pereira, G.H.}, \bibinfo{author}{Artes, R.}, \bibinfo{year}{2023}.
\newblock \bibinfo{title}{The circular quantile residual}.
\newblock \bibinfo{journal}{Computational Statistics \& Data Analysis} \bibinfo{volume}{178}, \bibinfo{pages}{107612}.
\bibitem[{Atkinson(1981)}]{atkinson1981two}
\bibinfo{author}{Atkinson, A.C.}, \bibinfo{year}{1981}.
\newblock \bibinfo{title}{Two graphical displays for outlying and influential observations in regression}.
\newblock \bibinfo{journal}{Biometrika} \bibinfo{volume}{68}, \bibinfo{pages}{13--20}.
\bibitem[{Breuninger et~al.(2021)Breuninger, Wawro, Breuninger, Reitmeier, Clavel, Six-Merker, Pestoni, Rohrmann, Rathmann, Peters et~al.}]{breuninger2021associations}
\bibinfo{author}{Breuninger, T.A.}, \bibinfo{author}{Wawro, N.}, \bibinfo{author}{Breuninger, J.}, \bibinfo{author}{Reitmeier, S.}, \bibinfo{author}{Clavel, T.}, \bibinfo{author}{Six-Merker, J.}, \bibinfo{author}{Pestoni, G.}, \bibinfo{author}{Rohrmann, S.}, \bibinfo{author}{Rathmann, W.}, \bibinfo{author}{Peters, A.}, et~al., \bibinfo{year}{2021}.
\newblock \bibinfo{title}{Associations between habitual diet, metabolic disease, and the gut microbiota using latent dirichlet allocation}.
\newblock \bibinfo{journal}{Microbiome} \bibinfo{volume}{9}, \bibinfo{pages}{1--18}.
\bibitem[{Chen et~al.(2020)Chen, Bernard, Padmapriya, Ning, Cai, Lan{\c{c}}a, Tan, Yap, Chong, Shek et~al.}]{chen2020associations}
\bibinfo{author}{Chen, B.}, \bibinfo{author}{Bernard, J.Y.}, \bibinfo{author}{Padmapriya, N.}, \bibinfo{author}{Ning, Y.}, \bibinfo{author}{Cai, S.}, \bibinfo{author}{Lan{\c{c}}a, C.}, \bibinfo{author}{Tan, K.H.}, \bibinfo{author}{Yap, F.}, \bibinfo{author}{Chong, Y.S.}, \bibinfo{author}{Shek, L.}, et~al., \bibinfo{year}{2020}.
\newblock \bibinfo{title}{Associations between early-life screen viewing and 24 hour movement behaviours: findings from a longitudinal birth cohort study}.
\newblock \bibinfo{journal}{The Lancet Child \& Adolescent Health} \bibinfo{volume}{4}, \bibinfo{pages}{201--209}.
\bibitem[{Da-Silva and Rodrigues(2015)}]{da2015bayesian}
\bibinfo{author}{Da-Silva, C.Q.}, \bibinfo{author}{Rodrigues, G.S.}, \bibinfo{year}{2015}.
\newblock \bibinfo{title}{Bayesian dynamic dirichlet models}.
\newblock \bibinfo{journal}{Communications in Statistics-Simulation and Computation} \bibinfo{volume}{44}, \bibinfo{pages}{787--818}.
\bibitem[{Das and Mukhopadhyay(2014)}]{das2014generalized}
\bibinfo{author}{Das, I.}, \bibinfo{author}{Mukhopadhyay, S.}, \bibinfo{year}{2014}.
\newblock \bibinfo{title}{On generalized multinomial models and joint percentile estimation}.
\newblock \bibinfo{journal}{Journal of Statistical Planning and Inference} \bibinfo{volume}{145}, \bibinfo{pages}{190--203}.
\bibitem[{Dunn and Smyth(1996)}]{dunn1996randomized}
\bibinfo{author}{Dunn, P.K.}, \bibinfo{author}{Smyth, G.K.}, \bibinfo{year}{1996}.
\newblock \bibinfo{title}{Randomized quantile residuals}.
\newblock \bibinfo{journal}{Journal of Computational and Graphical Statistics} \bibinfo{volume}{5}, \bibinfo{pages}{236--244}.
\bibitem[{Efron and Tibshirani(1994)}]{efron1994introduction}
\bibinfo{author}{Efron, B.}, \bibinfo{author}{Tibshirani, R.J.}, \bibinfo{year}{1994}.
\newblock \bibinfo{title}{An introduction to the bootstrap}.
\newblock \bibinfo{publisher}{CRC press}.
\bibitem[{Feng et~al.(2020)Feng, Li and Sadeghpour}]{feng2020comparison}
\bibinfo{author}{Feng, C.}, \bibinfo{author}{Li, L.}, \bibinfo{author}{Sadeghpour, A.}, \bibinfo{year}{2020}.
\newblock \bibinfo{title}{A comparison of residual diagnosis tools for diagnosing regression models for count data}.
\newblock \bibinfo{journal}{BMC Medical Research Methodology} \bibinfo{volume}{20}, \bibinfo{pages}{1--21}.
\bibitem[{Fernandes et~al.(2014)Fernandes, Reid, Macklaim, McMurrough, Edgell and Gloor}]{fernandes2014unifying}
\bibinfo{author}{Fernandes, A.D.}, \bibinfo{author}{Reid, J.N.}, \bibinfo{author}{Macklaim, J.M.}, \bibinfo{author}{McMurrough, T.A.}, \bibinfo{author}{Edgell, D.R.}, \bibinfo{author}{Gloor, G.B.}, \bibinfo{year}{2014}.
\newblock \bibinfo{title}{Unifying the analysis of high-throughput sequencing datasets: characterizing rna-seq, 16s rrna gene sequencing and selective growth experiments by compositional data analysis}.
\newblock \bibinfo{journal}{Microbiome} \bibinfo{volume}{2}, \bibinfo{pages}{1--13}.
\bibitem[{Ferrari and Cribari-Neto(2004)}]{ferrari2004beta}
\bibinfo{author}{Ferrari, S.}, \bibinfo{author}{Cribari-Neto, F.}, \bibinfo{year}{2004}.
\newblock \bibinfo{title}{Beta regression for modelling rates and proportions}.
\newblock \bibinfo{journal}{Journal of applied statistics} \bibinfo{volume}{31}, \bibinfo{pages}{799--815}.
\bibitem[{Filzmoser et~al.(2012)Filzmoser, Hron and Reimann}]{filzmoser2012interpretation}
\bibinfo{author}{Filzmoser, P.}, \bibinfo{author}{Hron, K.}, \bibinfo{author}{Reimann, C.}, \bibinfo{year}{2012}.
\newblock \bibinfo{title}{Interpretation of multivariate outliers for compositional data}.
\newblock \bibinfo{journal}{Computers \& Geosciences} \bibinfo{volume}{39}, \bibinfo{pages}{77--85}.
\bibitem[{Gonz{\'a}lez-Naranjo et~al.(2019)Gonz{\'a}lez-Naranjo, Alfonso-Alfonso, Grass-Fernandez, Morales-Chac{\'o}n, Pedroso-Ib{\'a}{\~n}ez, Ricardo-De La~Fe and Padr{\'o}n-S{\'a}nchez}]{gonzalez2019analysis}
\bibinfo{author}{Gonz{\'a}lez-Naranjo, J.E.}, \bibinfo{author}{Alfonso-Alfonso, M.}, \bibinfo{author}{Grass-Fernandez, D.}, \bibinfo{author}{Morales-Chac{\'o}n, L.M.}, \bibinfo{author}{Pedroso-Ib{\'a}{\~n}ez, I.}, \bibinfo{author}{Ricardo-De La~Fe, Y.}, \bibinfo{author}{Padr{\'o}n-S{\'a}nchez, A.}, \bibinfo{year}{2019}.
\newblock \bibinfo{title}{Analysis of sleep macrostructure in patients diagnosed with parkinson’s disease}.
\newblock \bibinfo{journal}{Behavioral Sciences} \bibinfo{volume}{9}, \bibinfo{pages}{6}.
\bibitem[{Gueorguieva et~al.(2008)Gueorguieva, Rosenheck and Zelterman}]{gueorguieva2008dirichlet}
\bibinfo{author}{Gueorguieva, R.}, \bibinfo{author}{Rosenheck, R.}, \bibinfo{author}{Zelterman, D.}, \bibinfo{year}{2008}.
\newblock \bibinfo{title}{Dirichlet component regression and its applications to psychiatric data}.
\newblock \bibinfo{journal}{Computational statistics \& data analysis} \bibinfo{volume}{52}, \bibinfo{pages}{5344--5355}.
\bibitem[{Hanretty(2021)}]{hanretty2021forecasting}
\bibinfo{author}{Hanretty, C.}, \bibinfo{year}{2021}.
\newblock \bibinfo{title}{Forecasting multiparty by-elections using dirichlet regression}.
\newblock \bibinfo{journal}{International Journal of Forecasting} \bibinfo{volume}{37}, \bibinfo{pages}{1666--1676}.
\bibitem[{Hijazi(2011)}]{hijazi2011algorithm}
\bibinfo{author}{Hijazi, R.}, \bibinfo{year}{2011}.
\newblock \bibinfo{title}{An em-algorithm based method to deal with rounded zeros in compositional data under dirichlet models}, in: \bibinfo{booktitle}{Proceedings of CoDaWork'11: 4th international workshop on Compositional Data Analysis, Egozcue, JJ, Tolosana-Delgado, R. and Ortego, MI (eds.) 2011}, \bibinfo{organization}{CIMNE}. pp. \bibinfo{pages}{1--10}.
\bibitem[{Hijazi(2006)}]{hijazi2006residuals}
\bibinfo{author}{Hijazi, R.H.}, \bibinfo{year}{2006}.
\newblock \bibinfo{title}{Residuals and diagnostics in dirichlet regression}.
\newblock \bibinfo{journal}{ASA Proceedings of the General Methodology Section} , \bibinfo{pages}{1190--1196}.
\bibitem[{Hijazi and Jernigan(2009)}]{hijazi2009modelling}
\bibinfo{author}{Hijazi, R.H.}, \bibinfo{author}{Jernigan, R.W.}, \bibinfo{year}{2009}.
\newblock \bibinfo{title}{Modelling compositional data using dirichlet regression models}.
\newblock \bibinfo{journal}{Journal of Applied Probability \& Statistics} \bibinfo{volume}{4}, \bibinfo{pages}{77--91}.
\bibitem[{Hosmer~Jr et~al.(2013)Hosmer~Jr, Lemeshow and Sturdivant}]{hosmer2013applied}
\bibinfo{author}{Hosmer~Jr, D.W.}, \bibinfo{author}{Lemeshow, S.}, \bibinfo{author}{Sturdivant, R.X.}, \bibinfo{year}{2013}.
\newblock \bibinfo{title}{Applied logistic regression}.
\newblock \bibinfo{edition}{3} ed., \bibinfo{publisher}{John Wiley \& Sons}.
\bibitem[{Hussain et~al.(2022)Hussain, Hossain, Jany, Bari, Uddin, Kamal, Ku and Kim}]{hussain2022quantitative}
\bibinfo{author}{Hussain, I.}, \bibinfo{author}{Hossain, M.A.}, \bibinfo{author}{Jany, R.}, \bibinfo{author}{Bari, M.A.}, \bibinfo{author}{Uddin, M.}, \bibinfo{author}{Kamal, A.R.M.}, \bibinfo{author}{Ku, Y.}, \bibinfo{author}{Kim, J.S.}, \bibinfo{year}{2022}.
\newblock \bibinfo{title}{Quantitative evaluation of eeg-biomarkers for prediction of sleep stages}.
\newblock \bibinfo{journal}{Sensors} \bibinfo{volume}{22}, \bibinfo{pages}{3079}.
\bibitem[{Johnson and Wichern(2013)}]{johnson2013applied}
\bibinfo{author}{Johnson, R.A.}, \bibinfo{author}{Wichern, D.W.}, \bibinfo{year}{2013}.
\newblock \bibinfo{title}{Applied multivariate statistical analysis}.
\newblock \bibinfo{edition}{6} ed., \bibinfo{publisher}{Prentice hall Upper Saddle River, NJ}.
\bibitem[{Lemonte and Moreno-Arenas(2019)}]{lemonte2019residuals}
\bibinfo{author}{Lemonte, A.J.}, \bibinfo{author}{Moreno-Arenas, G.}, \bibinfo{year}{2019}.
\newblock \bibinfo{title}{On residuals in generalized johnson sb regressions}.
\newblock \bibinfo{journal}{Applied Mathematical Modelling} \bibinfo{volume}{67}, \bibinfo{pages}{62--73}.
\bibitem[{Lemonte and Patriota(2011)}]{lemonte2011multivariate}
\bibinfo{author}{Lemonte, A.J.}, \bibinfo{author}{Patriota, A.G.}, \bibinfo{year}{2011}.
\newblock \bibinfo{title}{Multivariate elliptical models with general parameterization}.
\newblock \bibinfo{journal}{Statistical Methodology} \bibinfo{volume}{8}, \bibinfo{pages}{389--400}.
\bibitem[{Li et~al.(2022)Li, Al-Mahamda, Song, Feng and Sze}]{li2022alternate}
\bibinfo{author}{Li, D.}, \bibinfo{author}{Al-Mahamda, M.F.}, \bibinfo{author}{Song, Y.}, \bibinfo{author}{Feng, S.}, \bibinfo{author}{Sze, N.N.}, \bibinfo{year}{2022}.
\newblock \bibinfo{title}{An alternate crash severity multicategory modeling approach with asymmetric property}.
\newblock \bibinfo{journal}{Analytic Methods in Accident Research} \bibinfo{volume}{35}, \bibinfo{pages}{100218}.
\bibitem[{Maier(2014)}]{maier2014dirichletreg}
\bibinfo{author}{Maier, M.J.}, \bibinfo{year}{2014}.
\newblock \bibinfo{title}{Dirichletreg: Dirichlet regression for compositional data in r}.
\newblock \bibinfo{journal}{Research Report Series / Department of Statistics and Mathematics} \bibinfo{volume}{125}.
\bibitem[{Maski et~al.(2021)Maski, Colclasure, Little, Steinhart, Scammell, Navidi and Diniz~Behn}]{maski2021stability}
\bibinfo{author}{Maski, K.P.}, \bibinfo{author}{Colclasure, A.}, \bibinfo{author}{Little, E.}, \bibinfo{author}{Steinhart, E.}, \bibinfo{author}{Scammell, T.E.}, \bibinfo{author}{Navidi, W.}, \bibinfo{author}{Diniz~Behn, C.}, \bibinfo{year}{2021}.
\newblock \bibinfo{title}{Stability of nocturnal wake and sleep stages defines central nervous system disorders of hypersomnolence}.
\newblock \bibinfo{journal}{Sleep} \bibinfo{volume}{44}, \bibinfo{pages}{zsab021}.
\bibitem[{Melo et~al.(2022)Melo, Vargas, Lemonte and Moreno-Arenas}]{melo2022higher}
\bibinfo{author}{Melo, T.F.}, \bibinfo{author}{Vargas, T.M.}, \bibinfo{author}{Lemonte, A.J.}, \bibinfo{author}{Moreno-Arenas, G.}, \bibinfo{year}{2022}.
\newblock \bibinfo{title}{Higher-order asymptotic refinements in the multivariate dirichlet regression model}.
\newblock \bibinfo{journal}{Communications in Statistics-Simulation and Computation} \bibinfo{volume}{51}, \bibinfo{pages}{53--71}.
\bibitem[{Morais et~al.(2018)Morais, Thomas-Agnan and Simioni}]{morais2018using}
\bibinfo{author}{Morais, J.}, \bibinfo{author}{Thomas-Agnan, C.}, \bibinfo{author}{Simioni, M.}, \bibinfo{year}{2018}.
\newblock \bibinfo{title}{Using compositional and dirichlet models for market share regression}.
\newblock \bibinfo{journal}{Journal of Applied Statistics} \bibinfo{volume}{45}, \bibinfo{pages}{1670--1689}.
\bibitem[{Ohayon et~al.(2017)Ohayon, Wickwire, Hirshkowitz, Albert, Avidan, Daly, Dauvilliers, Ferri, Fung, Gozal et~al.}]{ohayon2017national}
\bibinfo{author}{Ohayon, M.}, \bibinfo{author}{Wickwire, E.M.}, \bibinfo{author}{Hirshkowitz, M.}, \bibinfo{author}{Albert, S.M.}, \bibinfo{author}{Avidan, A.}, \bibinfo{author}{Daly, F.J.}, \bibinfo{author}{Dauvilliers, Y.}, \bibinfo{author}{Ferri, R.}, \bibinfo{author}{Fung, C.}, \bibinfo{author}{Gozal, D.}, et~al., \bibinfo{year}{2017}.
\newblock \bibinfo{title}{National sleep foundation's sleep quality recommendations: first report}.
\newblock \bibinfo{journal}{Sleep health} \bibinfo{volume}{3}, \bibinfo{pages}{6--19}.
\bibitem[{Pereira et~al.(2020)Pereira, Scudilio, Santos-Neto, Botter and Sandoval}]{pereira2020class}
\bibinfo{author}{Pereira, G.H.}, \bibinfo{author}{Scudilio, J.}, \bibinfo{author}{Santos-Neto, M.}, \bibinfo{author}{Botter, D.A.}, \bibinfo{author}{Sandoval, M.C.}, \bibinfo{year}{2020}.
\newblock \bibinfo{title}{A class of residuals for outlier identification in zero adjusted regression models}.
\newblock \bibinfo{journal}{Journal of Applied Statistics} \bibinfo{volume}{47}, \bibinfo{pages}{1833--1847}.
\bibitem[{Pereira(2019)}]{pereira2019quantile}
\bibinfo{author}{Pereira, G.H.A.}, \bibinfo{year}{2019}.
\newblock \bibinfo{title}{On quantile residuals in beta regression}.
\newblock \bibinfo{journal}{Communications in Statistics-Simulation and Computation} \bibinfo{volume}{48}, \bibinfo{pages}{302--316}.
\bibitem[{Reichert et~al.(2022)Reichert, Berger, Dos~Santos, Lahdenper{\"a}, Nyein, Htut and Lummaa}]{reichert2022age}
\bibinfo{author}{Reichert, S.}, \bibinfo{author}{Berger, V.}, \bibinfo{author}{Dos~Santos, D.J.F.}, \bibinfo{author}{Lahdenper{\"a}, M.}, \bibinfo{author}{Nyein, U.K.}, \bibinfo{author}{Htut, W.}, \bibinfo{author}{Lummaa, V.}, \bibinfo{year}{2022}.
\newblock \bibinfo{title}{Age related variation of health markers in asian elephants}.
\newblock \bibinfo{journal}{Experimental Gerontology} \bibinfo{volume}{157}, \bibinfo{pages}{111629}.
\bibitem[{Ritmala-Castren et~al.(2015)Ritmala-Castren, Virtanen, Leivo, Kaukonen and Leino-Kilpi}]{ritmala2015sleep}
\bibinfo{author}{Ritmala-Castren, M.}, \bibinfo{author}{Virtanen, I.}, \bibinfo{author}{Leivo, S.}, \bibinfo{author}{Kaukonen, K.M.}, \bibinfo{author}{Leino-Kilpi, H.}, \bibinfo{year}{2015}.
\newblock \bibinfo{title}{Sleep and nursing care activities in an intensive care unit}.
\newblock \bibinfo{journal}{Nursing \& health sciences} \bibinfo{volume}{17}, \bibinfo{pages}{354--361}.
\bibitem[{Ross(2020)}]{ross2020tracking}
\bibinfo{author}{Ross, G.J.}, \bibinfo{year}{2020}.
\newblock \bibinfo{title}{Tracking the evolution of literary style via dirichlet--multinomial change point regression}.
\newblock \bibinfo{journal}{Journal of the Royal Statistical Society: Series A (Statistics in Society)} \bibinfo{volume}{183}, \bibinfo{pages}{149--167}.
\bibitem[{Schaid et~al.(2019)Schaid, Tong, Batzler, Sinnwell, Qing and Biernacka}]{schaid2019multivariate}
\bibinfo{author}{Schaid, D.J.}, \bibinfo{author}{Tong, X.}, \bibinfo{author}{Batzler, A.}, \bibinfo{author}{Sinnwell, J.P.}, \bibinfo{author}{Qing, J.}, \bibinfo{author}{Biernacka, J.M.}, \bibinfo{year}{2019}.
\newblock \bibinfo{title}{Multivariate generalized linear model for genetic pleiotropy}.
\newblock \bibinfo{journal}{Biostatistics} \bibinfo{volume}{20}, \bibinfo{pages}{111--128}.
\bibitem[{Scudilio and Pereira(2020)}]{scudilio2020adjusted}
\bibinfo{author}{Scudilio, J.}, \bibinfo{author}{Pereira, G.H.}, \bibinfo{year}{2020}.
\newblock \bibinfo{title}{Adjusted quantile residual for generalized linear models}.
\newblock \bibinfo{journal}{Computational Statistics} \bibinfo{volume}{35}, \bibinfo{pages}{399--421}.
\bibitem[{Sen et~al.(2010)Sen, Singer and de~Lima}]{sen2010finite}
\bibinfo{author}{Sen, P.K.}, \bibinfo{author}{Singer, J.M.}, \bibinfo{author}{de~Lima, A.C.P.}, \bibinfo{year}{2010}.
\newblock \bibinfo{title}{From finite sample to asymptotic methods in statistics}. volume~\bibinfo{volume}{28}.
\newblock \bibinfo{publisher}{Cambridge University Press}.
\bibitem[{Stephens(1986)}]{stephens1986tests}
\bibinfo{author}{Stephens, M.A.}, \bibinfo{year}{1986}.
\newblock \bibinfo{title}{Tests based on edf statistic}, in: \bibinfo{booktitle}{Goodness-of-fit Techniques}. \bibinfo{publisher}{Marcel Dekker}, pp. \bibinfo{pages}{97--193}.
\bibitem[{Tempesta et~al.(2018)Tempesta, Socci, De~Gennaro and Ferrara}]{tempesta2018sleep}
\bibinfo{author}{Tempesta, D.}, \bibinfo{author}{Socci, V.}, \bibinfo{author}{De~Gennaro, L.}, \bibinfo{author}{Ferrara, M.}, \bibinfo{year}{2018}.
\newblock \bibinfo{title}{Sleep and emotional processing}.
\newblock \bibinfo{journal}{Sleep medicine reviews} \bibinfo{volume}{40}, \bibinfo{pages}{183--195}.
\bibitem[{Tsagris and Stewart(2018)}]{tsagris2018dirichlet}
\bibinfo{author}{Tsagris, M.}, \bibinfo{author}{Stewart, C.}, \bibinfo{year}{2018}.
\newblock \bibinfo{title}{A dirichlet regression model for compositional data with zeros}.
\newblock \bibinfo{journal}{Lobachevskii Journal of Mathematics} \bibinfo{volume}{39}, \bibinfo{pages}{398--412}.
\bibitem[{Tsilimigras and Fodor(2016)}]{tsilimigras2016compositional}
\bibinfo{author}{Tsilimigras, M.C.}, \bibinfo{author}{Fodor, A.A.}, \bibinfo{year}{2016}.
\newblock \bibinfo{title}{Compositional data analysis of the microbiome: fundamentals, tools, and challenges}.
\newblock \bibinfo{journal}{Annals of epidemiology} \bibinfo{volume}{26}, \bibinfo{pages}{330--335}.
\bibitem[{Veje et~al.(2021)Veje, Studahl, Thunstr{\"o}m, Stentoft, Nolskog, Celik and Peker}]{veje2021sleep}
\bibinfo{author}{Veje, M.}, \bibinfo{author}{Studahl, M.}, \bibinfo{author}{Thunstr{\"o}m, E.}, \bibinfo{author}{Stentoft, E.}, \bibinfo{author}{Nolskog, P.}, \bibinfo{author}{Celik, Y.}, \bibinfo{author}{Peker, Y.}, \bibinfo{year}{2021}.
\newblock \bibinfo{title}{Sleep architecture, obstructive sleep apnea and functional outcomes in adults with a history of tick-borne encephalitis}.
\newblock \bibinfo{journal}{PLoS One} \bibinfo{volume}{16}, \bibinfo{pages}{e0246767}.
\bibitem[{Vieira and Salmon(2019)}]{vieira2019principled}
\bibinfo{author}{Vieira, B.H.}, \bibinfo{author}{Salmon, C.E.G.}, \bibinfo{year}{2019}.
\newblock \bibinfo{title}{A principled multivariate intersubject analysis of generalized partial directed coherence with dirichlet regression: Application to healthy aging in areas exhibiting cortical thinning}.
\newblock \bibinfo{journal}{Journal of Neuroscience Methods} \bibinfo{volume}{311}, \bibinfo{pages}{243--252}.
\bibitem[{Yoo et~al.(2022)Yoo, Sun, Ma, Chung and Kim}]{yoo2022guideline}
\bibinfo{author}{Yoo, J.}, \bibinfo{author}{Sun, Z.}, \bibinfo{author}{Ma, Q.}, \bibinfo{author}{Chung, D.}, \bibinfo{author}{Kim, Y.M.}, \bibinfo{year}{2022}.
\newblock \bibinfo{title}{A guideline for the statistical analysis of compositional data in immunology}.
\newblock \bibinfo{journal}{arXiv preprint arXiv:2201.07945} .
\bibitem[{Zhao et~al.(2019)Zhao, Westfall, Coulston, Lynch, Bullock and Montes}]{zhao2019additive}
\bibinfo{author}{Zhao, D.}, \bibinfo{author}{Westfall, J.}, \bibinfo{author}{Coulston, J.W.}, \bibinfo{author}{Lynch, T.B.}, \bibinfo{author}{Bullock, B.P.}, \bibinfo{author}{Montes, C.R.}, \bibinfo{year}{2019}.
\newblock \bibinfo{title}{Additive biomass equations for slash pine trees: Comparing three modeling approaches}.
\newblock \bibinfo{journal}{Canadian Journal of Forest Research} \bibinfo{volume}{49}, \bibinfo{pages}{27--40}.

\end{thebibliography}

\section*{Appendix}
\label{sec:appdx.demo}
\qquad
In this Appendix, we prove Theorem \ref{teo:asympt} using the definitions presented in Section \ref{sec:prop}. Under the correct model, if $\tau_i$ is consistently estimated, $y_i^{\{b\}}$, for $b=1,2,\hdots,B$, and $y_i$ have the same asymptotic distribution. As a result $l_i^{\{b\}}$ and $l_i$ have also the same asymptotic distribution and
\begin{equation}
\label{eq:aid}
a_i \xrightarrow{D} U_D(0;B),    
\end{equation}
where $U_D(0;B)$ denotes the discrete uniform distribution with parameters 0 and $B$. 

\vspace{0.3cm}

If $0 < k < 1/(B+1)$, then
$$
P(u_i \leq k) = P(u_i \leq k,a_i=0) =  P(u_i \leq k | a_i=0)P(a_i=0)=P(U^{\{1\}} \leq k)P(a_i=0),  
$$
where $U^{\{1\}}$ is a uniform distributed random variable with parameters 0 and $1/(B+1)$. Therefore, considering (\ref{eq:aid}),
\begin{equation}
\label{eq:ui1}
P(u_i \leq k) \xrightarrow{P} \frac{k}{1/(B+1)} \, \frac{1}{B+1}=k.  
\end{equation}

If $1/(B+1) \leq k < 1$, then
$$
P(u_i \leq k) = \sum_{j=0}^{\lambda-1}P(a_i=j) + P\left(\frac{\lambda}{B+1} < u_i \leq k,a_i=\lambda\right),  
$$
where $\lambda$ is the largest integer less or equal to $k(B+1)$. Then,
\begin{equation*}
\begin{split}
P(u_i \leq k) &= \sum_{j=0}^{\lambda-1}P(a_i=j) + 
P\left(\frac{\lambda}{B+1} < u_i \leq k | a_i=\lambda\right)P(a_i=\lambda) \\
&= \sum_{j=0}^{\lambda-1} P(a_i=j) + P(U^{\{2\}} \leq k)P(a_i=\lambda),
\end{split}
\end{equation*}
where $U^{\{2\}}$ is a uniform distributed random variable with parameters $\lambda/(B+1)$ and $(\lambda+1)/(B+1)$. Therefore, considering (\ref{eq:aid}),
\begin{equation}
\label{eq:ui2}
P(u_i \leq k) \xrightarrow{P} \frac{\lambda}{B+1} + \frac{k - \lambda/(B+1)}{1/(B+1)} \, \frac{1}{B+1} = \frac{\lambda}{B+1} + \frac{k(B+1)-\lambda}{B+1} = k.  
\end{equation}

From (\ref{eq:ui1}) and (\ref{eq:ui2}), $P(u_i \leq k) \xrightarrow{P} k$, for all $0 < k < 1$ and so
\begin{equation}
\label{eq:ui3}
u_i \xrightarrow{D} U(0;1). 
\end{equation}

Using (\ref{eq:ui3}) and the Sverdrup's theorem \citep[page 180]{sen2010finite},
$$
\Phi^{-1}\{u_i\}
\xrightarrow{D} \Phi^{-1}\{U(0;1)\},
$$ 
and so
$$
s_i \xrightarrow{D} N(0;1).
$$


\end{document}